\documentclass[%
preprint,
superscriptaddress,
showpacs,preprintnumbers,
bibnotes,
 amsmath,amssymb,
 aps,
pre,
]{revtex4-1}

\usepackage[T1]{fontenc}
\usepackage{mathptmx}
\usepackage{natbib}
\usepackage{float}
\usepackage{graphicx}
\usepackage{dcolumn}
\usepackage{subfigure}
\usepackage{epsfig}
\usepackage{epstopdf}
\usepackage{bm}
\usepackage[mathlines]{lineno}

\begin{document}

\newcommand{\reffig}[1]{Fig.\ref{#1}}
\newcommand{\equref}[1]{Eq. (\ref{#1})}

\newcommand{\cmm}{cm$^{-1}$ }
\newcommand{\ms}{MoS$_{2}$ }
\newcommand{\ag}{A$_{1g}$ }
\newcommand{\au}{A$_{2u}$ }
\newcommand{\eg}{E$_{2g}^1$ }
\newcommand{\nol}{number of layers }


\title{Resonant Raman imaging of MoS$_{2}$-substrate interaction}

\author{Hongyuan Li}
\affiliation{Institute for Quantum Science and Engineering, Texas A$\&$M University, College Station, TX, 77843 USA}
\affiliation{Department of Applied Physics, School of Science, Xi'an Jiaotong University, Shaanxi 710049, P. R. China}
\author{Dmitri V. Voronine}
\affiliation{Institute for Quantum Science and Engineering, Texas A$\&$M University, College Station, TX, 77843 USA}
\affiliation{Baylor University, Waco, TX, 76198 USA}

\begin{abstract}
We report a study of long-range \ms -substrate interaction using resonant Raman imaging. We observed a strong thickness-dependent peak shift of a Raman-forbidden mode that can be used as a new method of determining the thickness of multilayered \ms flakes. In addition, dependence of the Raman scattering intensity on thickness is explained by the interference enhancement theory. Finally, the resonant Raman spectra on different substrates are analysed. 
\end{abstract}

\maketitle

\section{Introduction}
Two-dimensional (2D) transition metal dichalcogenide (TMD) materials have recently attracted wide attention due to their potential applications in optoelectronic devices. For example, monolayer \ms has a large direct band gap \cite{p11}, and can be used in field-effect-transistors\cite{p12}. As a powerful tool, Raman spectroscopy has been widely used for studying the various properies of MoS$_2$. Frey $\it et. at.$ studied the resonant Raman (RR) spectra of \ms nanoparticles \cite{p6}. Li and Chakraborty have investigated the thickness-dependent effects for the Raman scattering of \ms \cite{paper1,li2012}. Also, the influence of the substrate-\ms interactions on the Raman spectra has been investigated \cite{p8}. In this paper, we performed RR imaging of multilayered \ms flakes using 660 nm laser excitation which corresponds to the A exciton of \ms\cite{p13,p14}. We obtained simultaneously the topographic information and correlated it with Raman imaging. The correlated AFM-Raman imaging reveals the relation between the thickness and optical properties. We also analysed the influence of substrates with different dielectric properties on the RR spectra of \ms.

\section{Results}
\begin{figure}[H]
\begin{minipage}[t]{0.34\linewidth}
  \centering
  \includegraphics[width=1\linewidth]{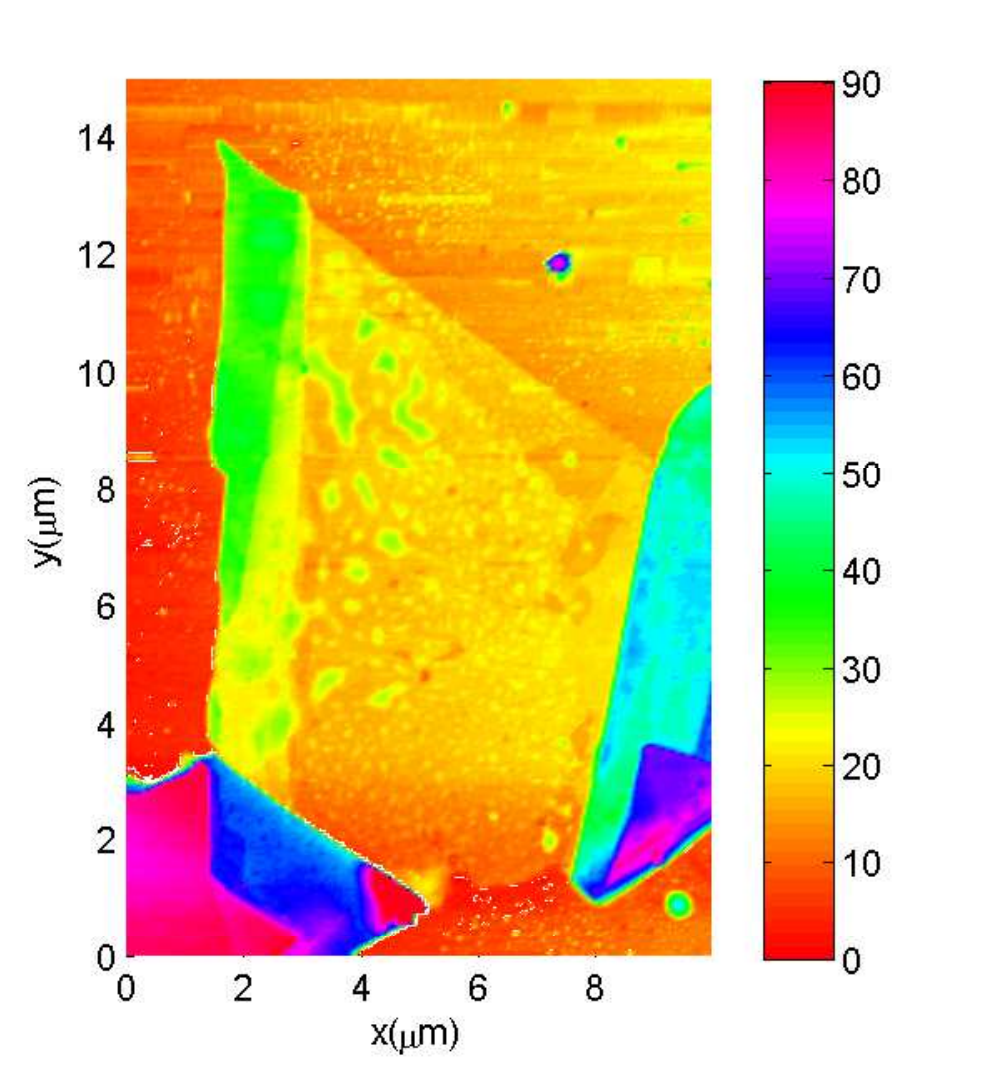}
      (\textbf{a})
\end{minipage}
\centering
\begin{minipage}[t]{0.65\linewidth}
  \centering
  \includegraphics[width=1\linewidth]{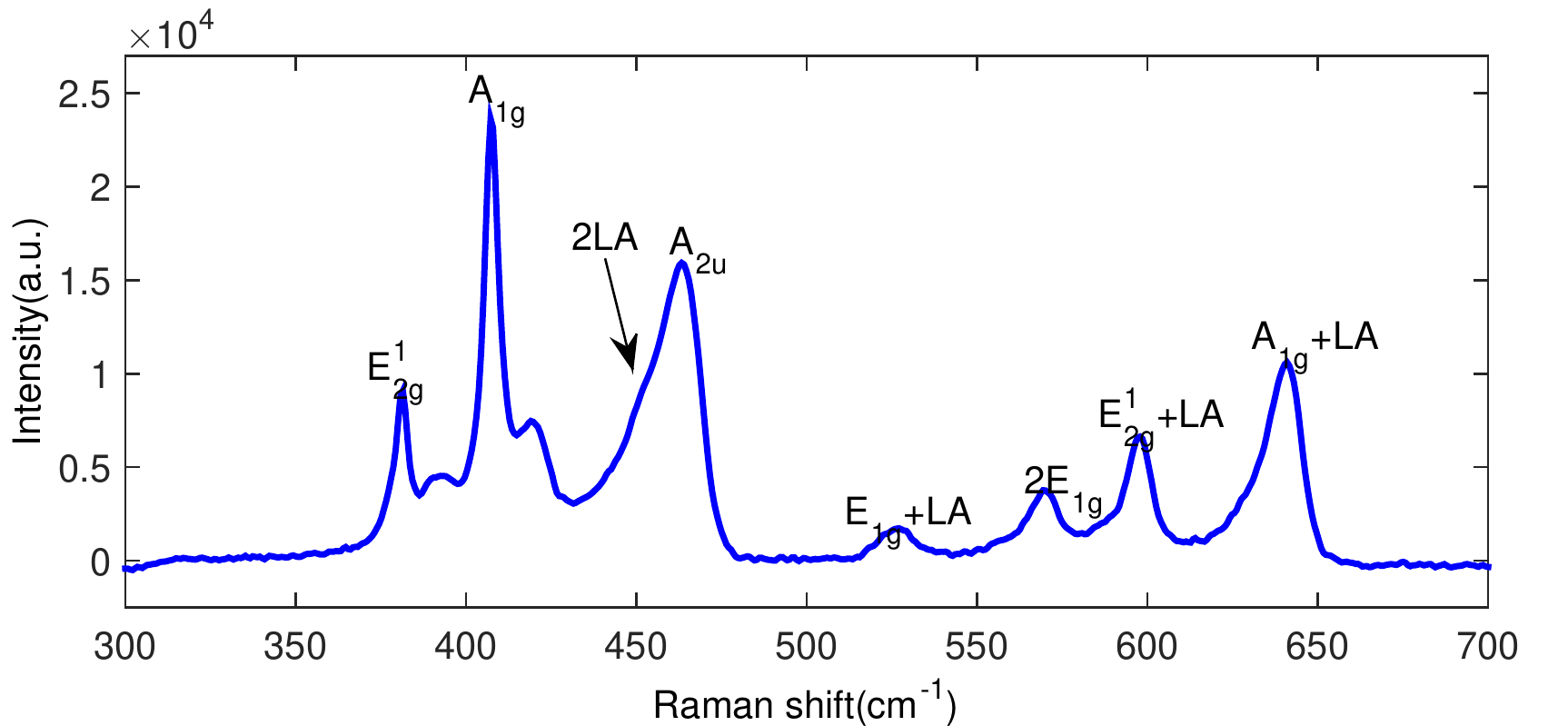}
      (\textbf{b})
\end{minipage}
\caption{AFM image (\textbf{a}) of \ms flake, with the colorbar representing the \nol and a typical resonant Raman spectrum (\textbf{b}) of a multilayered \ms flake on gold substrate excited by 660 nm laser.}\label{fig1}
\end{figure}

The multilayered \ms flakes were exfoliated on SiO$_2$ and gold substrate using the scotch-tape method. Th AFM image of a typical \ms flake is shown in \reffig{fig1} (\textbf{a}). The thinnest part of the flake has the height of $\sim$3nm, which corresponds to 5 layers of \ms \cite{paper1}. Note the colorbar in \reffig{fig1} (\textbf{a}) corresponds to the \nol.

A typical RR spectrum of the \ms flake is shown in \reffig{fig1} (\textbf{b}). The main spectral features include: strong out-of-plane \ag mode near 410\cmm, in-plane $E_{2g}^1$ mode near 385\cmm, IR-active \au mode near 466\cmm and several second-order modes, which include E$_{1g}$+LA at 528\cmm, 2E$_{1g}$ at 574\cmm, E$_{2g}^{1}$+LA at 600\cmm and A$_{1g}$+LA at 644\cmm \cite{paper1,p6}.

\begin{figure}
\begin{minipage}[t]{0.32\linewidth}
  \centering
  \includegraphics[width=1\textwidth]{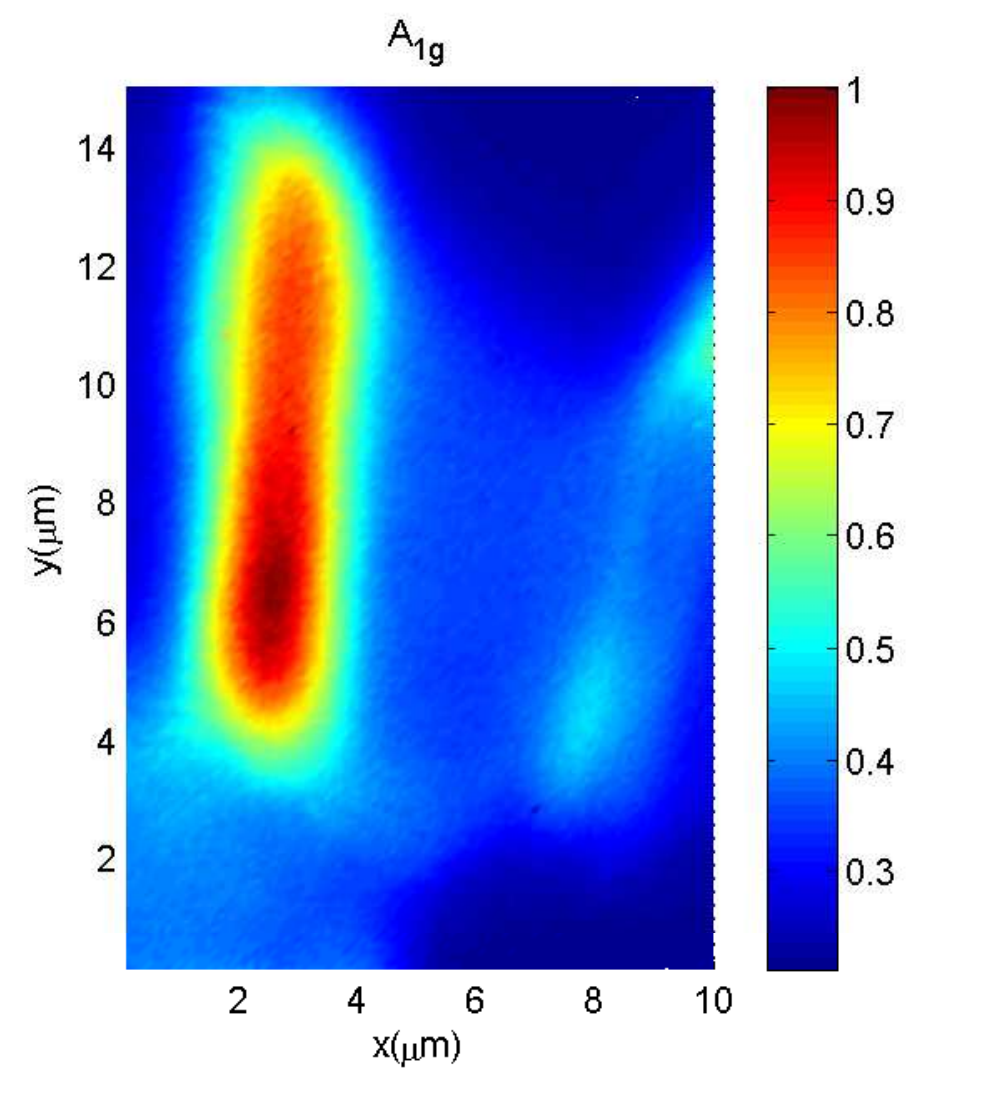}
    (\textbf{a})
\end{minipage}
\begin{minipage}[t]{0.32\linewidth}
    \centering
    \includegraphics[width=1\textwidth]{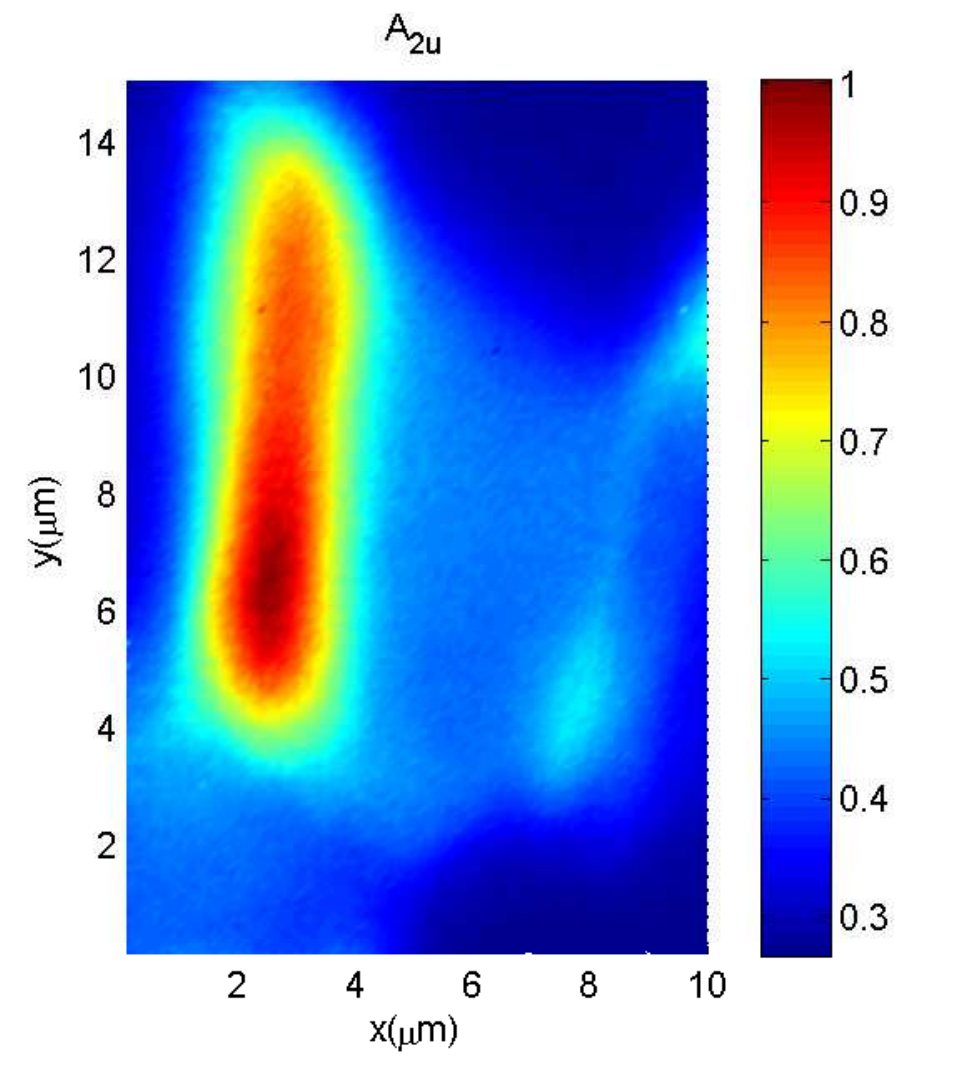}
    (\textbf{b})
\end{minipage}
\begin{minipage}[t]{0.32\linewidth}
    \centering
    \includegraphics[width=1\textwidth]{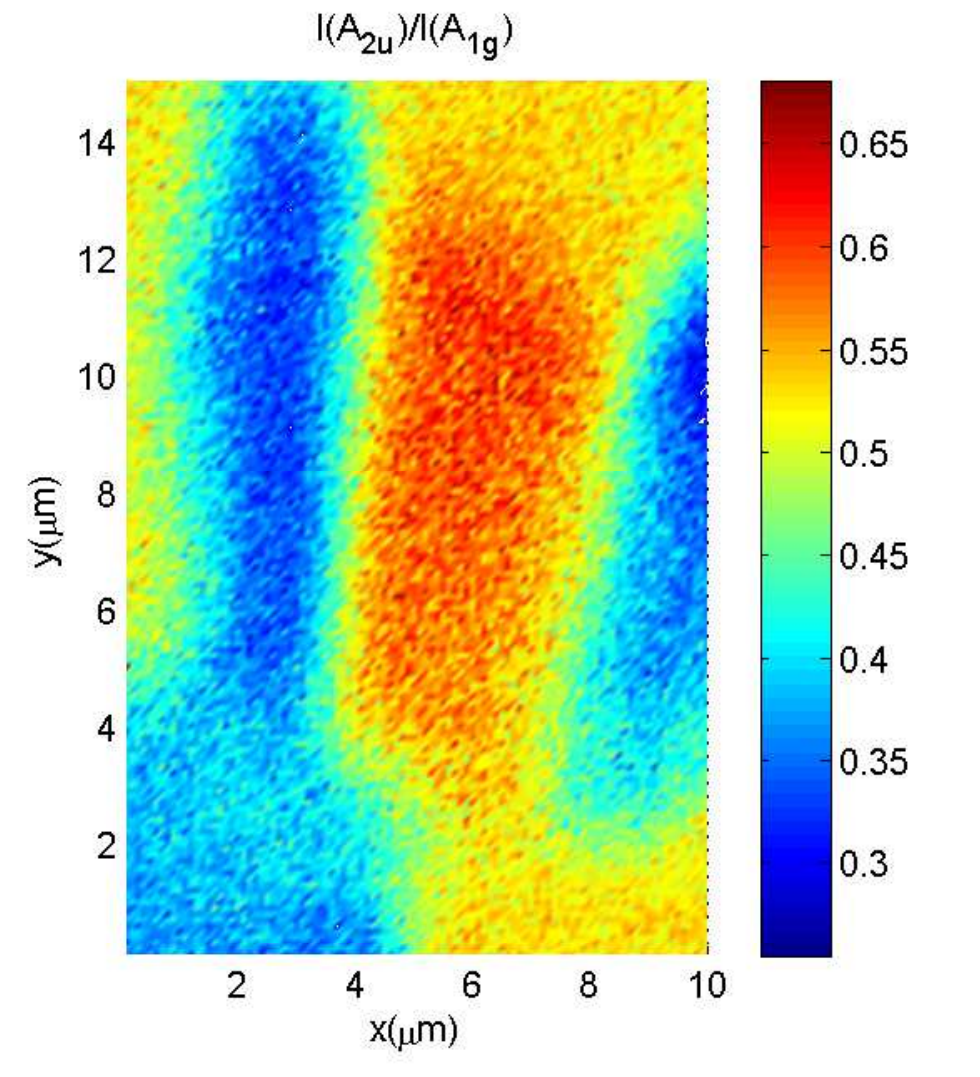}
    (\textbf{c})
\end{minipage}

\centering
\begin{minipage}[t]{0.32\linewidth}
    \centering
    \includegraphics[width=1\textwidth]{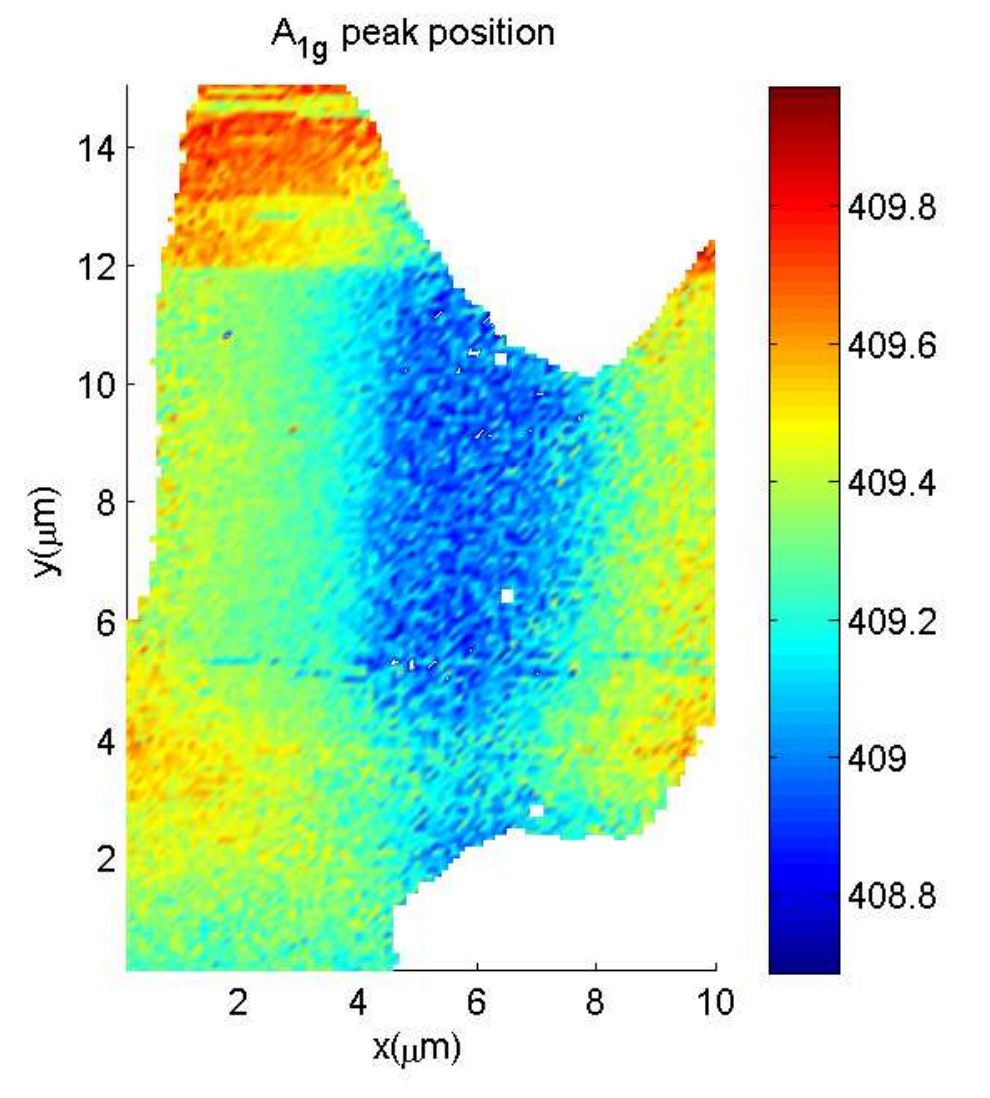}
    (\textbf{d})
\end{minipage}
\begin{minipage}[t]{0.32\linewidth}
    \centering
    \includegraphics[width=1\textwidth]{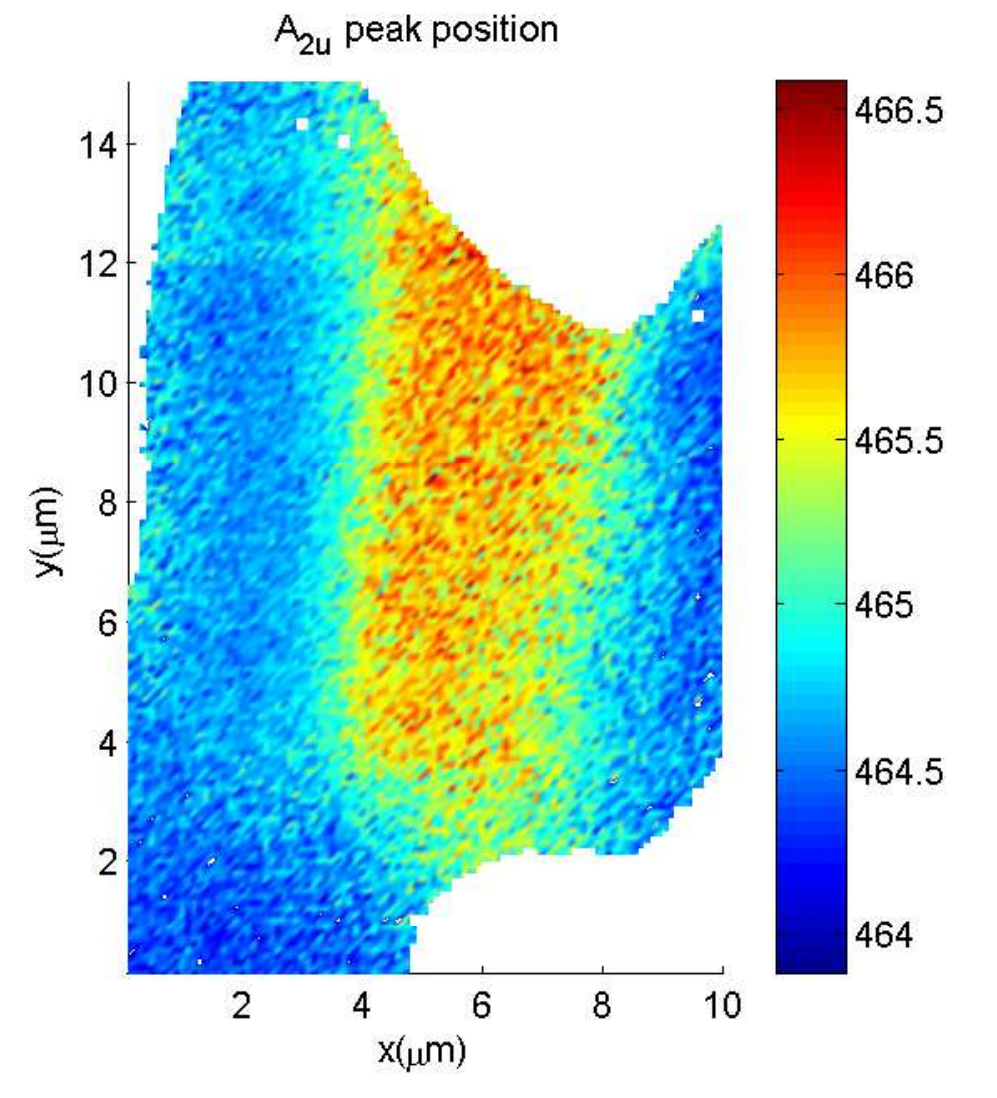}
    (\textbf{e})
\end{minipage}
\begin{minipage}[t]{0.32\linewidth}
    \centering
    \includegraphics[width=1\textwidth]{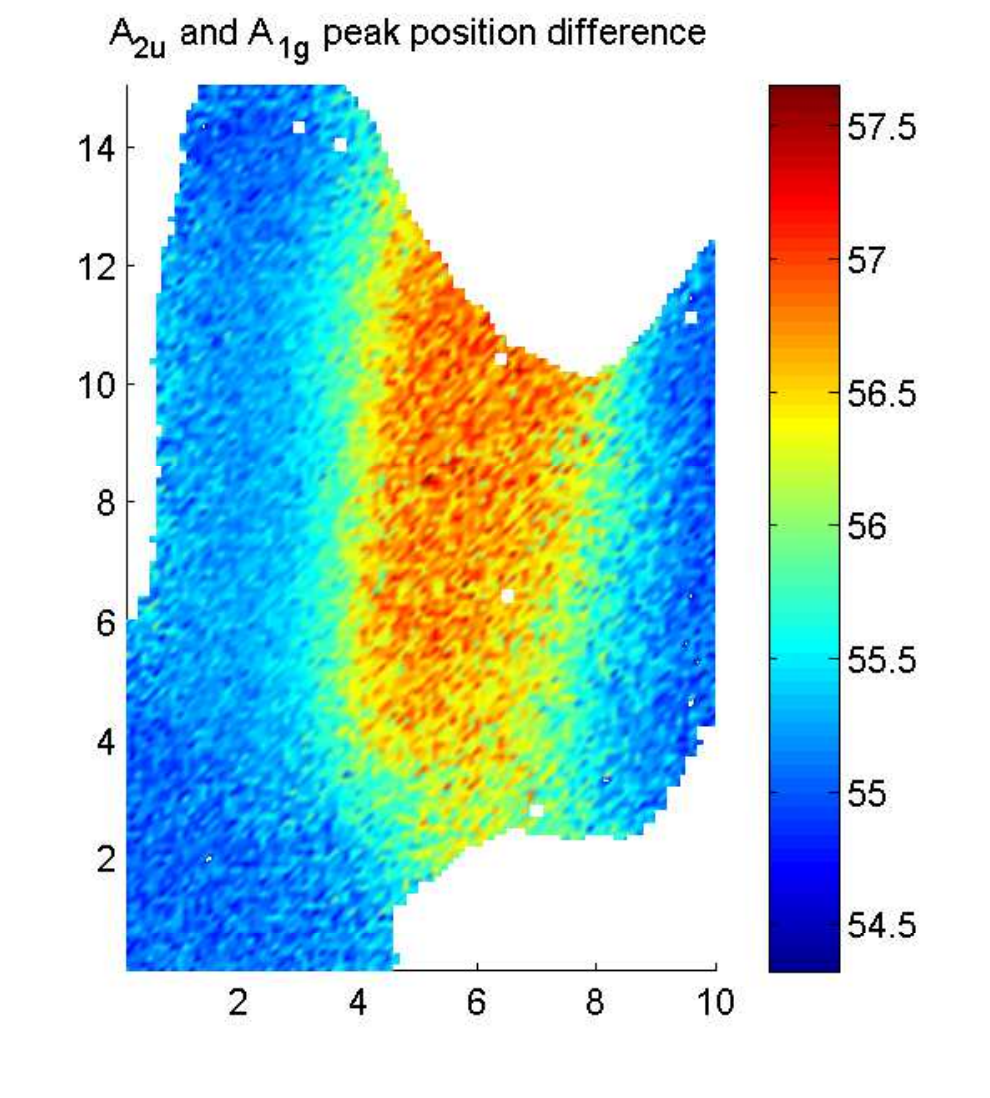}
    (\textbf{f})
\end{minipage}
\caption{Resonant Raman imaging of multilayered \ms on gold substrate: intensity of A$_{1g}$ (\textbf{a}) and A$_{2u}$ (\textbf{b}) modes, peak intensity ratio I(A$_{2u}$)/I({A$_{1g}$}) (\textbf{c}), peak portion maps for A$_{1g}$ (\textbf{d}) and A$_{2u}$ (\textbf{e}) modes, and peak position difference A$_{2u}$-A$_{1g}$ (\textbf{f}).}\label{fig3}
\end{figure}

Resonant Raman imaging of the multilayered \ms flake on the gold substrate is shown in \reffig{fig3}. The imaging step size was 100nm. The intensity maps for \ag and \au modes are shown in \reffig{fig3} (\textbf{a}) and (\textbf{b}). The intensity ratio I(\au)/I(\ag) is shown in \reffig{fig3} (\textbf{c}). Both \reffig{fig3} (\textbf{a}) and (\textbf{b}) show that the area with the strongest intensity corresponds to the \nol of about 30, while the very thin (eg: the middle) and the very thick (eg: the left bottom) parts of the flake show a relatively low intensity. I(\ag) vs the \nol is shown in \reffig{fig5} (\textbf{b}) where the maximum intensity correspond to n=30. The intensity ratios in \reffig{fig5} (\textbf{a}) and \reffig{fig6} indicate that for the thin part, the \au peak has a relatively stronger intensity compared to other modes.

\begin{figure}
\begin{minipage}[t]{0.49\linewidth}
    \includegraphics[width=1\textwidth]{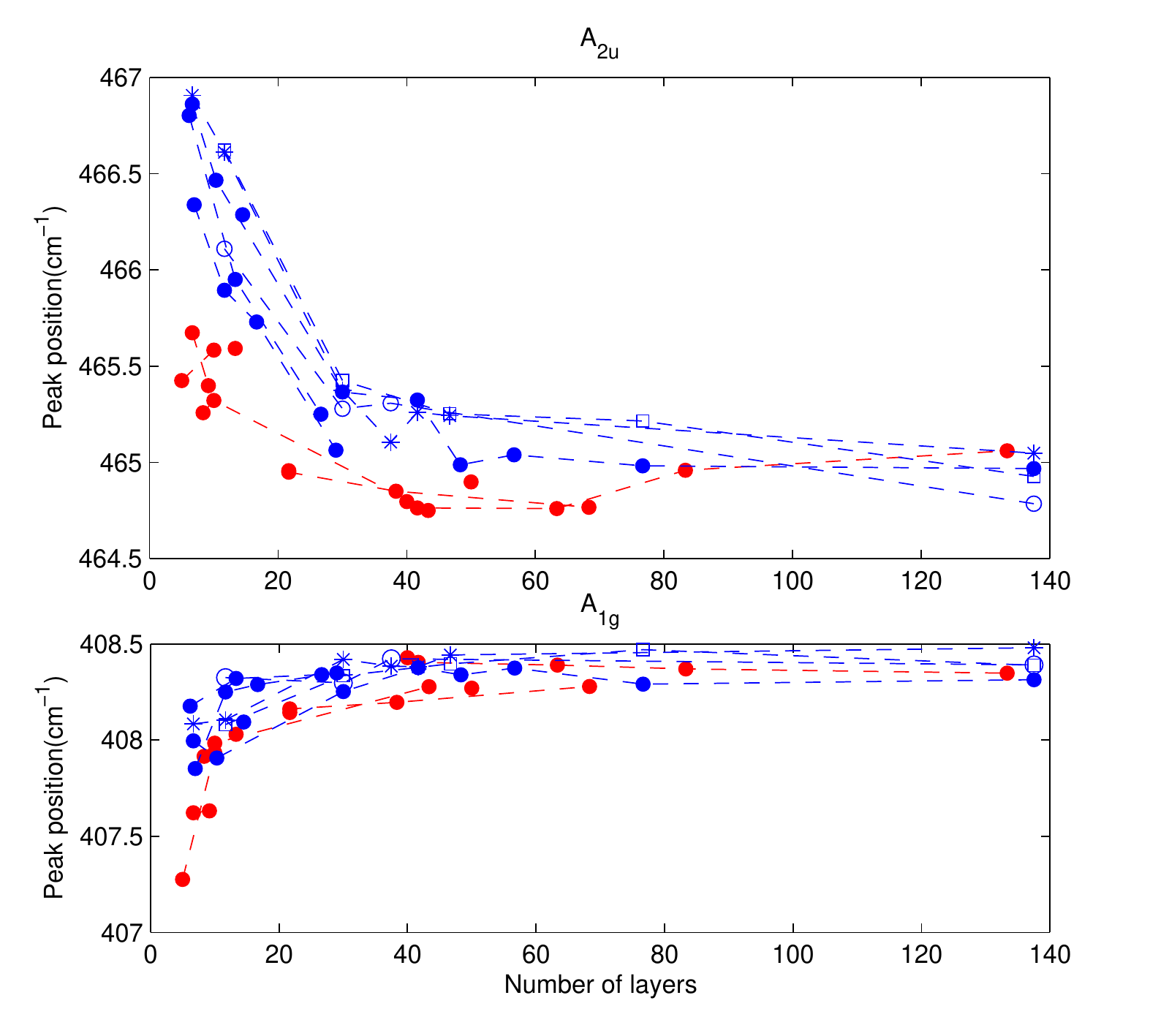}
    (\textbf{a})
\end{minipage}
\begin{minipage}[t]{0.49\linewidth}
    \includegraphics[width=1\textwidth]{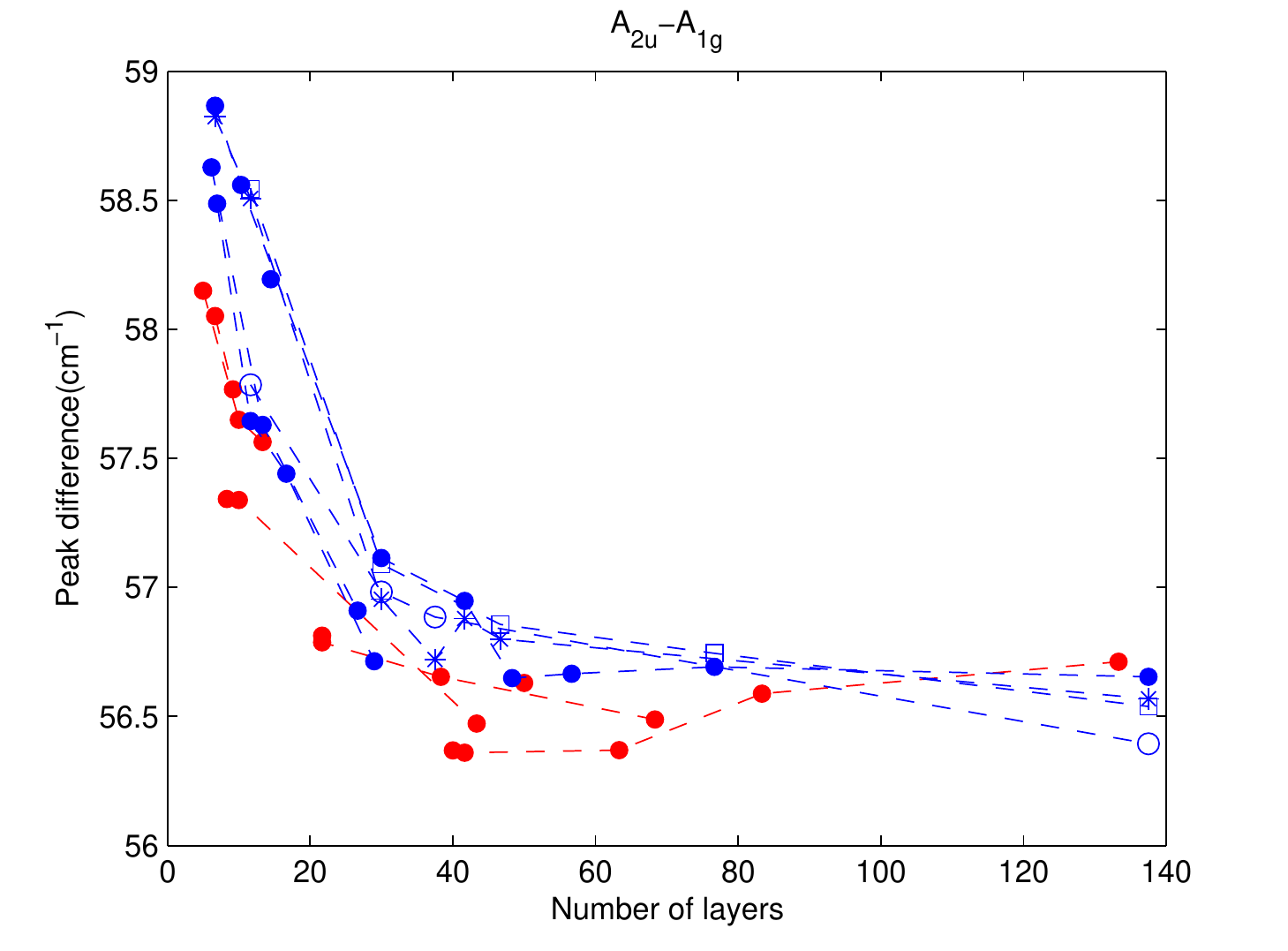}
    (\textbf{b})
\end{minipage}

\centering
\begin{minipage}[t]{0.49\linewidth}
    \centering
    \includegraphics[width=1\textwidth]{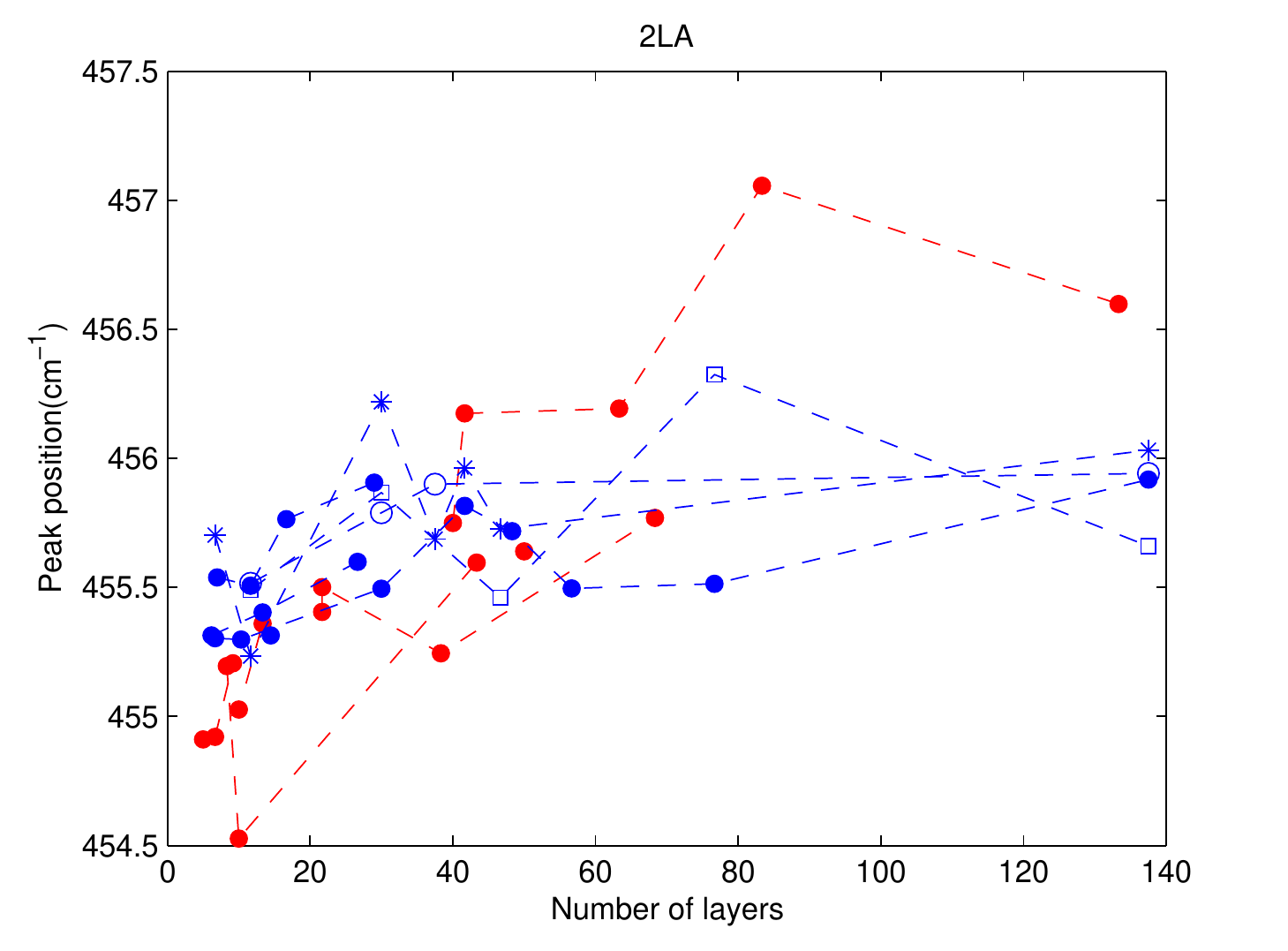}
    (\textbf{c})
\end{minipage}
\caption{Thickness-dependence of peak position. (\textbf{a}) Peak position of A$_{2u}$ and A$_{1g}$ modes. (\textbf{b}) Peak position difference A$_{2u}$-A$_{1g}$. (\textbf{c}) Peak position of 2LA. Blue and red dots represent the measurements on gold and SiO$_2$ substrates, respectively. Data from several flaskes are shown as groups of points linked by dashed lines with all connected points belonging to the same flake}\label{fig4}
\end{figure}

\begin{figure}
\begin{minipage}[t]{0.49\linewidth}
    \centering
    \includegraphics[width=1\textwidth]{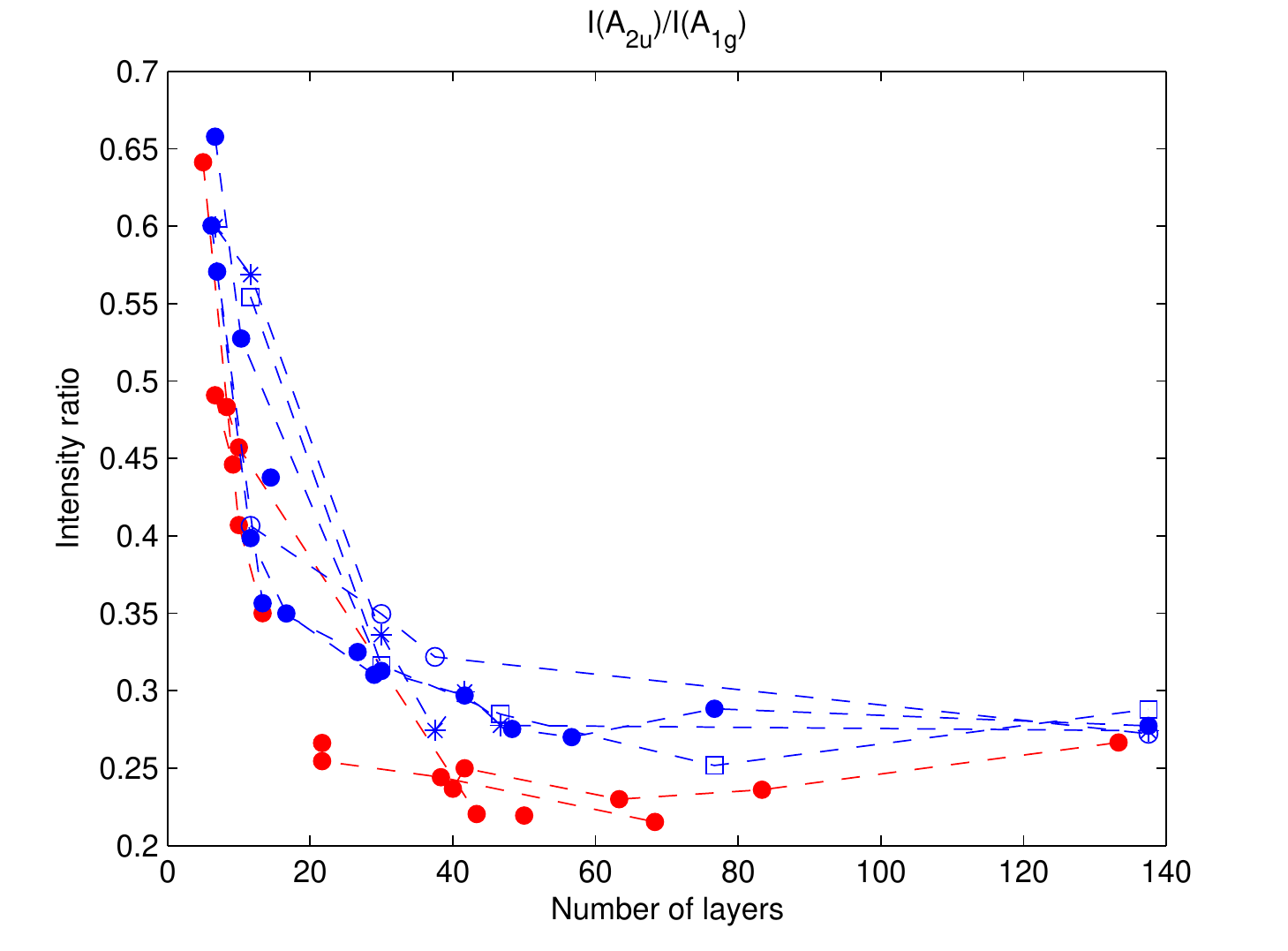}
    (\textbf{a})
\end{minipage}
\begin{minipage}[t]{0.49\linewidth}
    \centering
    \includegraphics[width=1\textwidth]{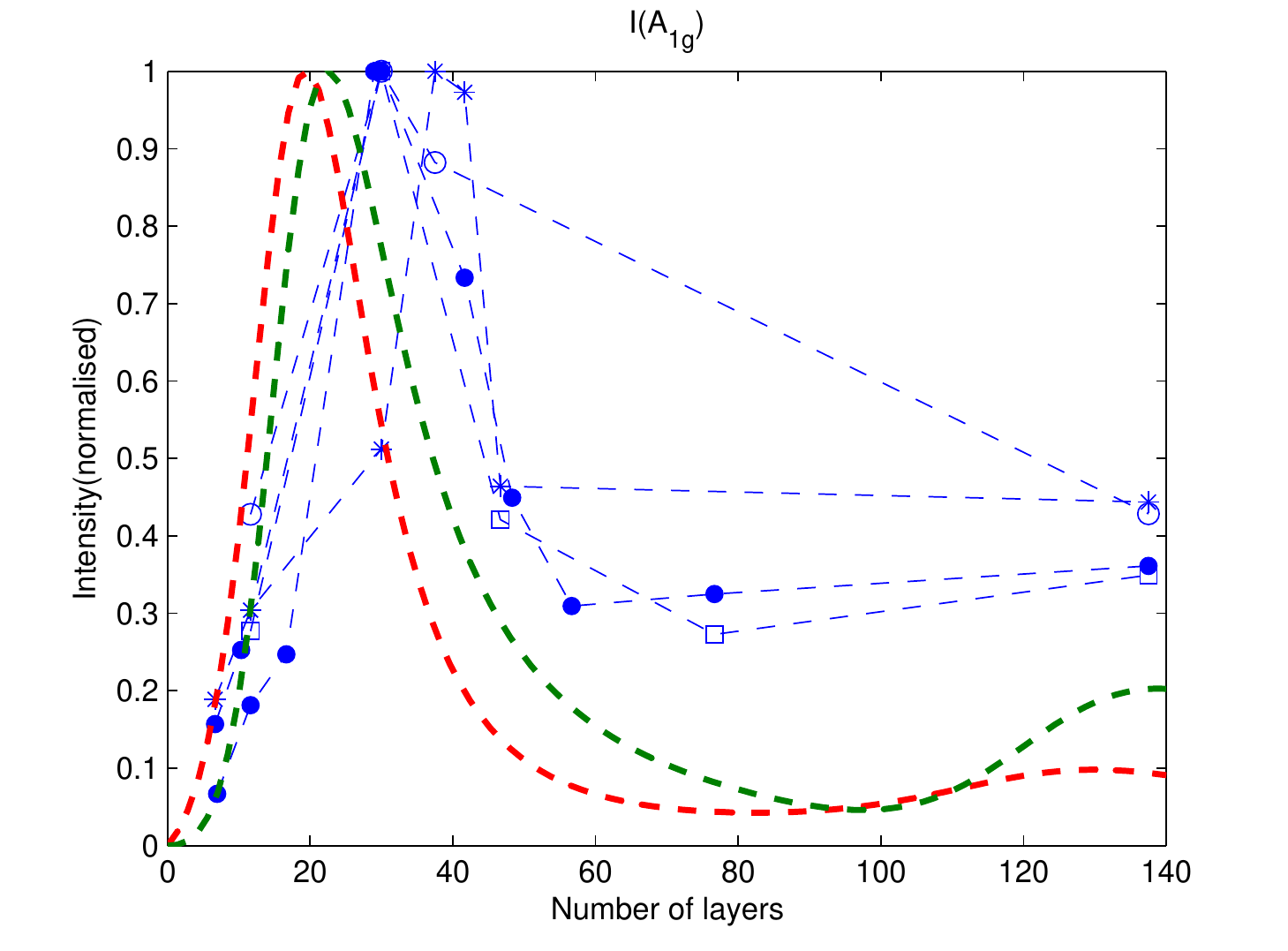}
    (\textbf{b})
\end{minipage}
\caption{(\textbf{a}).The intensity ratio I(A$_{2u}$)/I(A$_{1g}$) decreases rapidly with the increase of the thickness and stablizes after a turning point of $\sim$10nm. (\textbf{b}) A$_{1g}$ intensity as a function of thickness showes a maximum at $\sim$30 layers. Red dashed line shows the simulation}\label{fig5}
\end{figure}

\begin{figure}
\begin{minipage}[t]{0.49\linewidth}
    \centering
    \includegraphics[width=1\textwidth]{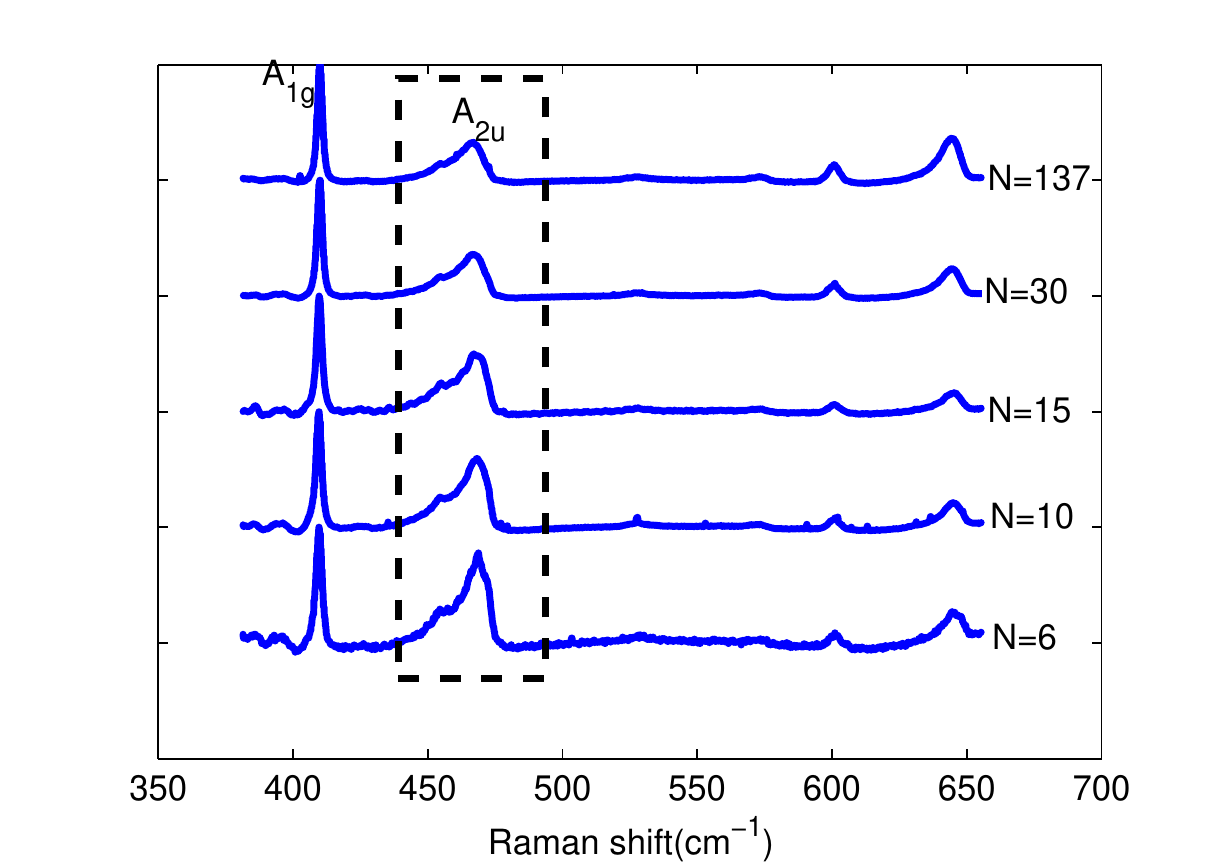}
    (\textbf{a})
\end{minipage}
\begin{minipage}[t]{0.49\linewidth}
    \centering
    \includegraphics[width=1\textwidth]{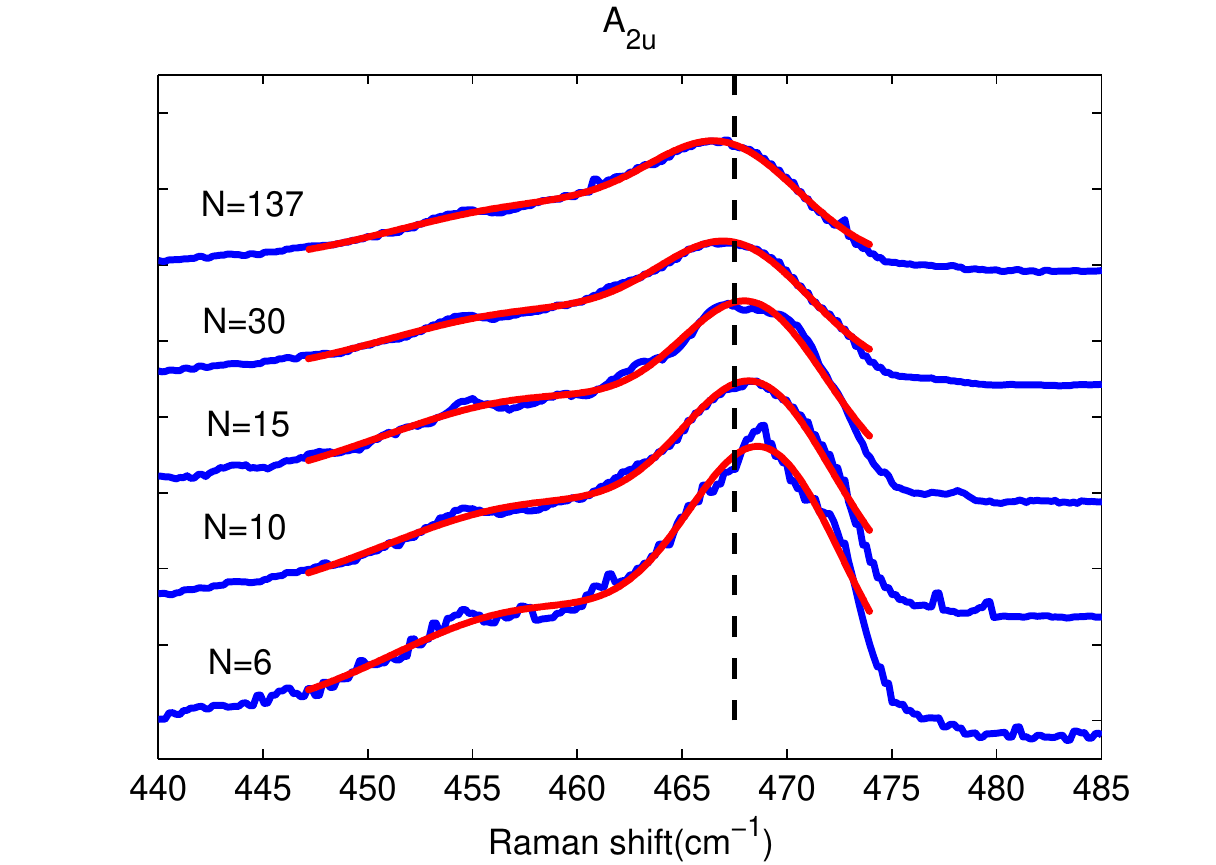}
    (\textbf{b})
\end{minipage}
\caption{(\textbf{a}) Resonant Raman spectra for different \nol N(normalised to A$_{1g}$ and offset for convenience). (\textbf{b}) Spectra of the A$_{2u}$ mode, corresponding to the dashed spectral region in (\textbf{a}).}\label{fig6}
\end{figure}

Peak positions also show shifts with the change of the \nol. The peak position maps of the \ag and \au modes are shown in \reffig{fig3} (\textbf{d}) and (\textbf{e}). Combined with the height distribution in \reffig{fig1} (\textbf{a}), the RR spectra show a blue shift for \ag and a red shift for \au for a multilayered flake compared to bulk \ms. \reffig{fig4} (\textbf{a}) and (\textbf{b}) show the \ag and \au peak position versus the \nol. The \ag peak position increases slightly($\sim$0.5\cmm) when the \nol increases from 7 to 140, while the \au peak has a blue shift of about $\sim$2\cmm. The shift of the \au peak can seen more clearly in \reffig{fig6} (\textbf{a}) and (\textbf{b}).
%

\section{Discussion}
We consider two main effects: (1) substrate dependence, (2) thickness-dependence, including peak shifts and intensity variations. It has been widely reported that, on common insulating substrates such as silicone and SiO$_2$, the increase of the \nol has an influence on the two characteristic Raman peaks E$_{2g}^1 $and \ag\cite{paper1,p2}. The \ag mode has a blue shift due to the increase of the force constant which is induced by the increased interlayer van der Waals interaction. Our RR measurements of the \ag peak position variation are consistent with the previous reports. However, here we show that, under the resonant condition, the Raman-inactive \au mode shows a strong thickness-dependent softening with the increase of the \nol. This is attributed to the decrease of the long-range Coulombic interaction between the Mo atoms with increasing layer number \cite{p3}. The results on SiO$_2$ substrate are consistent with previous reports.  For the small \nol(n$<$10), the thickness-dependent RR signal of both \ag and \au mode show red shifts with the increase of the \nol\cite{paper1}. When the \nol exceeds $\sim$10, the change of the peak position is small and difficult to measure.

However on the gold substrate, we observed a previously unreported blue shift for the \au mode. \reffig{fig4} (\textbf{a}) shows that for a small \nol($\sim6$), the \au peak has a red shift of about 2\cmm compared to the bulk \ms. With the increase of the thickness, this red shift gradually disappears with a turning point near $n=30$, as shown in \reffig{fig4} (\textbf{a}). This indicates that the thickness-dependent effects can be observed within a large range of the \nol. This is different from the SiO$_2$ substrate where the peak shift can only be observed for a small \nol(n$<$10).

We consider two possible reasons for the different peak shift directions of the \au mode of the multilayered \ms flake on the gold substrate: (1) laser-induced heating of the gold substrate may have thermal effects on \ms leading to the peak shift \cite{p7}; (2) charge transfer from \ms to the gold substrate modifying the doping, and the electron-phonon interactions \cite{p8,p9}. We discard the first possibility by performing laser power dependent measurements.  Different laser power can lead to different surface temperatures. However, we find no observable changes in the peak position vs the \nol for different laser powers (\reffig{fig4}). The second possible explanation may be supported by previous literature reports which considered the non-resonant situation. Several studies reported p-doping of \ms using various methods, including deposition on the gold substrate \cite{p8,p19}, decoration with gold nano-particles (NP) \cite{p17}, and using monolayer \ms transistor to adjust the doping level directly \cite{p18} resulting in stiffening of the \ag mode. Here our results show the corresponding blue shift of the \ag mode on the gold substrate for the number of layers N<10, which is consistent with the previous studies. Here, we, for the first time, observed that under the resonant conditions, the Raman-inactive \au mode of \ms on the gold substrate shows a strong blue shift compared with that on the SiO$_2$ substrate. \reffig{fig4} (\textbf{a}) shows a shift more than 1\cmm when the \nol is less than 10 layers.

\reffig{fig5} shows the nearly linear increase of the \ag peak intensity until the \nol reaches  $\sim$30 with the maximum point n=30, followed by a decrease until n=50. Note that the Raman signal intensity of bulk \ms is weaker than that with the \nol of $\sim$30. This may be attributed to the inference enhancement\cite{p4,p5} due to multiple reflections of the incident laser and emitted Raman signals in \ms flakes. Using the model of Wang $\it et. al.$, the intensity of the Raman signal can be expressed as
\begin{equation}\label{eq1}
I=\int_0^d |t\gamma|^2 dy,
\end{equation}
where $d$ is the thickness of the \ms flake, $t$ is the amplitude of the electric field at the depth $y$, and $\gamma$ is the enhancement factor due to the multiple-reflection. The simulation is shown as a red dashed line in \reffig{fig4} (\textbf{b}). The original model assumes no coherence between the field scattered from adjacent layers. Considering coherence, \equref{eq1} can be written as
\begin{equation}\label{eq2}
I=\left |\int_0^d t\gamma dy\right |^2.
\end{equation}
The coherent simulation is different, shown as a green dashed line in \reffig{fig4} (\textbf{b}). Zeng $\it et. al.$ provided a qualititative explanation by using the accumulated phase shift for the $n_{th}$ reflected field and destructive interference with the first reflected field from the flake surface \cite{p16}.

\section*{acknowledgements}
We thank Profs. Marlan Scully,  Alexei Sokolov and Zhenrong Zhang for helpful discussions. We also thank Prof. Marlan Scully for the access to Raman facilities and we thank Zhenrong Zhang for help with sample preparation. D.V. acknowledges the support of NSF grant CHE-1609608.

\bibliographystyle{apsrev4-1}
\bibliography{reference}

\begin{thebibliography}{18}%
\makeatletter
\providecommand \@ifxundefined [1]{%
 \@ifx{#1\undefined}
}%
\providecommand \@ifnum [1]{%
 \ifnum #1\expandafter \@firstoftwo
 \else \expandafter \@secondoftwo
 \fi
}%
\providecommand \@ifx [1]{%
 \ifx #1\expandafter \@firstoftwo
 \else \expandafter \@secondoftwo
 \fi
}%
\providecommand \natexlab [1]{#1}%
\providecommand \enquote  [1]{``#1''}%
\providecommand \bibnamefont  [1]{#1}%
\providecommand \bibfnamefont [1]{#1}%
\providecommand \citenamefont [1]{#1}%
\providecommand \href@noop [0]{\@secondoftwo}%
\providecommand \href [0]{\begingroup \@sanitize@url \@href}%
\providecommand \@href[1]{\@@startlink{#1}\@@href}%
\providecommand \@@href[1]{\endgroup#1\@@endlink}%
\providecommand \@sanitize@url [0]{\catcode `\\12\catcode `\$12\catcode
  `\&12\catcode `\#12\catcode `\^12\catcode `\_12\catcode `\%12\relax}%
\providecommand \@@startlink[1]{}%
\providecommand \@@endlink[0]{}%
\providecommand \url  [0]{\begingroup\@sanitize@url \@url }%
\providecommand \@url [1]{\endgroup\@href {#1}{\urlprefix }}%
\providecommand \urlprefix  [0]{URL }%
\providecommand \Eprint [0]{\href }%
\providecommand \doibase [0]{http://dx.doi.org/}%
\providecommand \selectlanguage [0]{\@gobble}%
\providecommand \bibinfo  [0]{\@secondoftwo}%
\providecommand \bibfield  [0]{\@secondoftwo}%
\providecommand \translation [1]{[#1]}%
\providecommand \BibitemOpen [0]{}%
\providecommand \bibitemStop [0]{}%
\providecommand \bibitemNoStop [0]{.\EOS\space}%
\providecommand \EOS [0]{\spacefactor3000\relax}%
\providecommand \BibitemShut  [1]{\csname bibitem#1\endcsname}%
\let\auto@bib@innerbib\@empty
\bibitem [{\citenamefont {Splendiani}\ \emph {et~al.}(2010)\citenamefont
  {Splendiani}, \citenamefont {Sun}, \citenamefont {Zhang}, \citenamefont {Li},
  \citenamefont {Kim}, \citenamefont {Chim}, \citenamefont {Galli},\ and\
  \citenamefont {Wang}}]{p11}%
  \BibitemOpen
  \bibfield  {author} {\bibinfo {author} {\bibfnamefont {A.}~\bibnamefont
  {Splendiani}}, \bibinfo {author} {\bibfnamefont {L.}~\bibnamefont {Sun}},
  \bibinfo {author} {\bibfnamefont {Y.~B.}\ \bibnamefont {Zhang}}, \bibinfo
  {author} {\bibfnamefont {T.~S.}\ \bibnamefont {Li}}, \bibinfo {author}
  {\bibfnamefont {J.}~\bibnamefont {Kim}}, \bibinfo {author} {\bibfnamefont
  {C.~Y.}\ \bibnamefont {Chim}}, \bibinfo {author} {\bibfnamefont
  {G.}~\bibnamefont {Galli}}, \ and\ \bibinfo {author} {\bibfnamefont
  {F.}~\bibnamefont {Wang}},\ }\href@noop {} {\bibfield  {journal} {\bibinfo
  {journal} {Nano Lett.}\ }\textbf {\bibinfo {volume} {10}},\ \bibinfo {pages}
  {1271} (\bibinfo {year} {2010})}\BibitemShut {NoStop}%
\bibitem [{\citenamefont {Lin}\ \emph {et~al.}(2013)\citenamefont {Lin},
  \citenamefont {Li}, \citenamefont {Zhang},\ and\ \citenamefont {Chen}}]{p12}%
  \BibitemOpen
  \bibfield  {author} {\bibinfo {author} {\bibfnamefont {J.}~\bibnamefont
  {Lin}}, \bibinfo {author} {\bibfnamefont {H.}~\bibnamefont {Li}}, \bibinfo
  {author} {\bibfnamefont {H.}~\bibnamefont {Zhang}}, \ and\ \bibinfo {author}
  {\bibfnamefont {W.}~\bibnamefont {Chen}},\ }\href@noop {} {\bibfield
  {journal} {\bibinfo  {journal} {Appl. Phys. Lett.}\ }\textbf {\bibinfo
  {volume} {102}},\ \bibinfo {pages} {203109} (\bibinfo {year}
  {2013})}\BibitemShut {NoStop}%
\bibitem [{\citenamefont {Frey}\ \emph {et~al.}(1999)\citenamefont {Frey},
  \citenamefont {Tenne}, \citenamefont {Matthews}, \citenamefont
  {Dresselhaus},\ and\ \citenamefont {Dresselhaus}}]{p6}%
  \BibitemOpen
  \bibfield  {author} {\bibinfo {author} {\bibfnamefont {G.~L.}\ \bibnamefont
  {Frey}}, \bibinfo {author} {\bibfnamefont {R.}~\bibnamefont {Tenne}},
  \bibinfo {author} {\bibfnamefont {M.~J.}\ \bibnamefont {Matthews}}, \bibinfo
  {author} {\bibfnamefont {M.}~\bibnamefont {Dresselhaus}}, \ and\ \bibinfo
  {author} {\bibfnamefont {G.}~\bibnamefont {Dresselhaus}},\ }\href@noop {}
  {\bibfield  {journal} {\bibinfo  {journal} {Physical Review B}\ }\textbf
  {\bibinfo {volume} {60}},\ \bibinfo {pages} {2883} (\bibinfo {year}
  {1999})}\BibitemShut {NoStop}%
\bibitem [{\citenamefont {Chakraborty}\ \emph {et~al.}(2013)\citenamefont
  {Chakraborty}, \citenamefont {Matte}, \citenamefont {Sood},\ and\
  \citenamefont {Rao}}]{paper1}%
  \BibitemOpen
  \bibfield  {author} {\bibinfo {author} {\bibfnamefont {B.}~\bibnamefont
  {Chakraborty}}, \bibinfo {author} {\bibfnamefont {H.~S. S.~R.}\ \bibnamefont
  {Matte}}, \bibinfo {author} {\bibfnamefont {A.~K.}\ \bibnamefont {Sood}}, \
  and\ \bibinfo {author} {\bibfnamefont {C.~N.~R.}\ \bibnamefont {Rao}},\
  }\href@noop {} {\bibfield  {journal} {\bibinfo  {journal} {J. Raman
  Spectrosc.}\ }\textbf {\bibinfo {volume} {44}},\ \bibinfo {pages} {92}
  (\bibinfo {year} {2013})}\BibitemShut {NoStop}%
\bibitem [{\citenamefont {Li}\ \emph {et~al.}(2012{\natexlab{a}})\citenamefont
  {Li}, \citenamefont {Zhang}, \citenamefont {Yap}, \citenamefont {Tay},
  \citenamefont {Edwin}, \citenamefont {Olivier},\ and\ \citenamefont
  {Baillargeat}}]{li2012}%
  \BibitemOpen
  \bibfield  {author} {\bibinfo {author} {\bibfnamefont {H.}~\bibnamefont
  {Li}}, \bibinfo {author} {\bibfnamefont {Q.}~\bibnamefont {Zhang}}, \bibinfo
  {author} {\bibfnamefont {C.~C.~R.}\ \bibnamefont {Yap}}, \bibinfo {author}
  {\bibfnamefont {B.~K.}\ \bibnamefont {Tay}}, \bibinfo {author} {\bibfnamefont
  {T.~H.~T.}\ \bibnamefont {Edwin}}, \bibinfo {author} {\bibfnamefont
  {A.}~\bibnamefont {Olivier}}, \ and\ \bibinfo {author} {\bibfnamefont
  {D.}~\bibnamefont {Baillargeat}},\ }\href@noop {} {\bibfield  {journal}
  {\bibinfo  {journal} {Advanced Functional Materials}\ }\textbf {\bibinfo
  {volume} {22}},\ \bibinfo {pages} {1385} (\bibinfo {year}
  {2012}{\natexlab{a}})}\BibitemShut {NoStop}%
\bibitem [{\citenamefont {Buscema}\ \emph {et~al.}(2014)\citenamefont
  {Buscema}, \citenamefont {Steele}, \citenamefont {van~der Zant},\ and\
  \citenamefont {Castellanos-Gomez}}]{p8}%
  \BibitemOpen
  \bibfield  {author} {\bibinfo {author} {\bibfnamefont {M.}~\bibnamefont
  {Buscema}}, \bibinfo {author} {\bibfnamefont {G.~A.}\ \bibnamefont {Steele}},
  \bibinfo {author} {\bibfnamefont {H.~S.~J.}\ \bibnamefont {van~der Zant}}, \
  and\ \bibinfo {author} {\bibfnamefont {A.}~\bibnamefont
  {Castellanos-Gomez}},\ }\href@noop {} {\bibfield  {journal} {\bibinfo
  {journal} {Nano Res.}\ }\textbf {\bibinfo {volume} {7}},\ \bibinfo {pages}
  {561} (\bibinfo {year} {2014})}\BibitemShut {NoStop}%
\bibitem [{\citenamefont {Coehoorn}\ \emph {et~al.}(1987)\citenamefont
  {Coehoorn}, \citenamefont {Hass},\ and\ \citenamefont {de~Groot}}]{p13}%
  \BibitemOpen
  \bibfield  {author} {\bibinfo {author} {\bibfnamefont {R.}~\bibnamefont
  {Coehoorn}}, \bibinfo {author} {\bibfnamefont {C.}~\bibnamefont {Hass}}, \
  and\ \bibinfo {author} {\bibfnamefont {R.~A.}\ \bibnamefont {de~Groot}},\
  }\href@noop {} {\bibfield  {journal} {\bibinfo  {journal} {Phys. Rev. B}\
  }\textbf {\bibinfo {volume} {35}},\ \bibinfo {pages} {6023} (\bibinfo {year}
  {1987})}\BibitemShut {NoStop}%
\bibitem [{\citenamefont {Acrivos}\ \emph {et~al.}(1971)\citenamefont
  {Acrivos}, \citenamefont {Liang}, \citenamefont {Wilson},\ and\ \citenamefont
  {Yoffe}}]{p14}%
  \BibitemOpen
  \bibfield  {author} {\bibinfo {author} {\bibfnamefont {J.~V.}\ \bibnamefont
  {Acrivos}}, \bibinfo {author} {\bibfnamefont {W.~Y.}\ \bibnamefont {Liang}},
  \bibinfo {author} {\bibfnamefont {J.~A.}\ \bibnamefont {Wilson}}, \ and\
  \bibinfo {author} {\bibfnamefont {A.~D.}\ \bibnamefont {Yoffe}},\ }\href@noop
  {} {\bibfield  {journal} {\bibinfo  {journal} {J. Phy. C}\ }\textbf {\bibinfo
  {volume} {4}},\ \bibinfo {pages} {L18} (\bibinfo {year} {1971})}\BibitemShut
  {NoStop}%
\bibitem [{\citenamefont {Li}\ \emph {et~al.}(2012{\natexlab{b}})\citenamefont
  {Li}, \citenamefont {Zhang}, \citenamefont {Yap}, \citenamefont {Tay},
  \citenamefont {Edwin}, \citenamefont {Olivier},\ and\ \citenamefont
  {Baillargeat}}]{p2}%
  \BibitemOpen
  \bibfield  {author} {\bibinfo {author} {\bibfnamefont {H.}~\bibnamefont
  {Li}}, \bibinfo {author} {\bibfnamefont {Q.}~\bibnamefont {Zhang}}, \bibinfo
  {author} {\bibfnamefont {C.~C.~R.}\ \bibnamefont {Yap}}, \bibinfo {author}
  {\bibfnamefont {B.~K.}\ \bibnamefont {Tay}}, \bibinfo {author} {\bibfnamefont
  {T.~H.~T.}\ \bibnamefont {Edwin}}, \bibinfo {author} {\bibfnamefont
  {A.}~\bibnamefont {Olivier}}, \ and\ \bibinfo {author} {\bibfnamefont
  {D.}~\bibnamefont {Baillargeat}},\ }\href@noop {} {\bibfield  {journal}
  {\bibinfo  {journal} {Adv. Funct. Mater.}\ }\textbf {\bibinfo {volume}
  {22}},\ \bibinfo {pages} {1385} (\bibinfo {year}
  {2012}{\natexlab{b}})}\BibitemShut {NoStop}%
\bibitem [{\citenamefont {Sanchez}\ and\ \citenamefont {Wirtz}(2011)}]{p3}%
  \BibitemOpen
  \bibfield  {author} {\bibinfo {author} {\bibfnamefont {A.~M.}\ \bibnamefont
  {Sanchez}}\ and\ \bibinfo {author} {\bibfnamefont {L.}~\bibnamefont
  {Wirtz}},\ }\href@noop {} {\bibfield  {journal} {\bibinfo  {journal} {Phys.
  Rev. B}\ }\textbf {\bibinfo {volume} {84}},\ \bibinfo {pages} {155413}
  (\bibinfo {year} {2011})}\BibitemShut {NoStop}%
\bibitem [{\citenamefont {Najmaei}\ \emph {et~al.}(2014)\citenamefont
  {Najmaei}, \citenamefont {Mlayah}, \citenamefont {Arbouet}, \citenamefont
  {Girard}, \citenamefont {Leotin},\ and\ \citenamefont {Lou}}]{p7}%
  \BibitemOpen
  \bibfield  {author} {\bibinfo {author} {\bibfnamefont {S.}~\bibnamefont
  {Najmaei}}, \bibinfo {author} {\bibfnamefont {A.}~\bibnamefont {Mlayah}},
  \bibinfo {author} {\bibfnamefont {A.}~\bibnamefont {Arbouet}}, \bibinfo
  {author} {\bibfnamefont {C.}~\bibnamefont {Girard}}, \bibinfo {author}
  {\bibfnamefont {J.}~\bibnamefont {Leotin}}, \ and\ \bibinfo {author}
  {\bibfnamefont {J.}~\bibnamefont {Lou}},\ }\href@noop {} {\bibfield
  {journal} {\bibinfo  {journal} {NANO}\ }\textbf {\bibinfo {volume} {8}},\
  \bibinfo {pages} {12682} (\bibinfo {year} {2014})}\BibitemShut {NoStop}%
\bibitem [{\citenamefont {Bhanu}\ \emph {et~al.}(2014)\citenamefont {Bhanu},
  \citenamefont {Islam}, \citenamefont {Tetard},\ and\ \citenamefont
  {Khondaker}}]{p9}%
  \BibitemOpen
  \bibfield  {author} {\bibinfo {author} {\bibfnamefont {U.}~\bibnamefont
  {Bhanu}}, \bibinfo {author} {\bibfnamefont {M.~R.}\ \bibnamefont {Islam}},
  \bibinfo {author} {\bibfnamefont {L.}~\bibnamefont {Tetard}}, \ and\ \bibinfo
  {author} {\bibfnamefont {S.~I.}\ \bibnamefont {Khondaker}},\ }\href@noop {}
  {\bibfield  {journal} {\bibinfo  {journal} {Sci. Rep.}\ }\textbf {\bibinfo
  {volume} {4}},\ \bibinfo {pages} {5574} (\bibinfo {year} {2014})}\BibitemShut
  {NoStop}%
\bibitem [{\citenamefont {Robinson}\ \emph {et~al.}(2015)\citenamefont
  {Robinson}, \citenamefont {Giusca}, \citenamefont {Gonzalez}, \citenamefont
  {Kay}, \citenamefont {Kazakova},\ and\ \citenamefont {Kolosov}}]{p19}%
  \BibitemOpen
  \bibfield  {author} {\bibinfo {author} {\bibfnamefont {B.~J.}\ \bibnamefont
  {Robinson}}, \bibinfo {author} {\bibfnamefont {C.~E.}\ \bibnamefont
  {Giusca}}, \bibinfo {author} {\bibfnamefont {Y.~T.}\ \bibnamefont
  {Gonzalez}}, \bibinfo {author} {\bibfnamefont {N.~D.}\ \bibnamefont {Kay}},
  \bibinfo {author} {\bibfnamefont {O.}~\bibnamefont {Kazakova}}, \ and\
  \bibinfo {author} {\bibfnamefont {O.~V.}\ \bibnamefont {Kolosov}},\
  }\href@noop {} {\bibfield  {journal} {\bibinfo  {journal} {2D Materials}\
  }\textbf {\bibinfo {volume} {2}},\ \bibinfo {pages} {015005} (\bibinfo {year}
  {2015})}\BibitemShut {NoStop}%
\bibitem [{\citenamefont {Y.Shi}\ \emph {et~al.}(2013)\citenamefont {Y.Shi},
  \citenamefont {Huang}, \citenamefont {Jin3}, \citenamefont {Hsu},
  \citenamefont {Yu}, \citenamefont {Li},\ and\ \citenamefont {Yang}}]{p17}%
  \BibitemOpen
  \bibfield  {author} {\bibinfo {author} {\bibnamefont {Y.Shi}}, \bibinfo
  {author} {\bibfnamefont {J.}~\bibnamefont {Huang}}, \bibinfo {author}
  {\bibfnamefont {L.}~\bibnamefont {Jin3}}, \bibinfo {author} {\bibfnamefont
  {Y.}~\bibnamefont {Hsu}}, \bibinfo {author} {\bibfnamefont {S.}~\bibnamefont
  {Yu}}, \bibinfo {author} {\bibfnamefont {L.}~\bibnamefont {Li}}, \ and\
  \bibinfo {author} {\bibfnamefont {H.}~\bibnamefont {Yang}},\ }\href@noop {}
  {\bibfield  {journal} {\bibinfo  {journal} {Scientific Reports}\ }\textbf
  {\bibinfo {volume} {3}},\ \bibinfo {pages} {1839} (\bibinfo {year}
  {2013})}\BibitemShut {NoStop}%
\bibitem [{\citenamefont {Chakraborty}\ \emph {et~al.}(2012)\citenamefont
  {Chakraborty}, \citenamefont {Bera}, \citenamefont {Muthu}, \citenamefont
  {Bhowmick}, \citenamefont {Waghmare},\ and\ \citenamefont {Sood}}]{p18}%
  \BibitemOpen
  \bibfield  {author} {\bibinfo {author} {\bibfnamefont {B.}~\bibnamefont
  {Chakraborty}}, \bibinfo {author} {\bibfnamefont {A.}~\bibnamefont {Bera}},
  \bibinfo {author} {\bibfnamefont {D.}~\bibnamefont {Muthu}}, \bibinfo
  {author} {\bibfnamefont {S.}~\bibnamefont {Bhowmick}}, \bibinfo {author}
  {\bibfnamefont {U.~V.}\ \bibnamefont {Waghmare}}, \ and\ \bibinfo {author}
  {\bibfnamefont {A.}~\bibnamefont {Sood}},\ }\href@noop {} {\bibfield
  {journal} {\bibinfo  {journal} {Phys. Rev. B}\ }\textbf {\bibinfo {volume}
  {85}},\ \bibinfo {pages} {161403} (\bibinfo {year} {2012})}\BibitemShut
  {NoStop}%
\bibitem [{\citenamefont {Wang}\ \emph {et~al.}(2008)\citenamefont {Wang},
  \citenamefont {Ni}, \citenamefont {Shen}, \citenamefont {Wang}, ,\ and\
  \citenamefont {Wu}}]{p4}%
  \BibitemOpen
  \bibfield  {author} {\bibinfo {author} {\bibfnamefont {Y.~Y.}\ \bibnamefont
  {Wang}}, \bibinfo {author} {\bibfnamefont {Z.~H.}\ \bibnamefont {Ni}},
  \bibinfo {author} {\bibfnamefont {Z.~X.}\ \bibnamefont {Shen}}, \bibinfo
  {author} {\bibfnamefont {H.~M.}\ \bibnamefont {Wang}}, , \ and\ \bibinfo
  {author} {\bibfnamefont {Y.~H.}\ \bibnamefont {Wu}},\ }\href@noop {}
  {\bibfield  {journal} {\bibinfo  {journal} {Appl. Phys. Lett.}\ }\textbf
  {\bibinfo {volume} {92}},\ \bibinfo {pages} {043121} (\bibinfo {year}
  {2008})}\BibitemShut {NoStop}%
\bibitem [{\citenamefont {Ni}\ \emph {et~al.}(2008)\citenamefont {Ni},
  \citenamefont {Wang}, \citenamefont {Yu},\ and\ \citenamefont {Shen}}]{p5}%
  \BibitemOpen
  \bibfield  {author} {\bibinfo {author} {\bibfnamefont {Z.}~\bibnamefont
  {Ni}}, \bibinfo {author} {\bibfnamefont {Y.}~\bibnamefont {Wang}}, \bibinfo
  {author} {\bibfnamefont {T.}~\bibnamefont {Yu}}, \ and\ \bibinfo {author}
  {\bibfnamefont {Z.}~\bibnamefont {Shen}},\ }\href@noop {} {\bibfield
  {journal} {\bibinfo  {journal} {Nano Res}\ }\textbf {\bibinfo {volume} {1}},\
  \bibinfo {pages} {273} (\bibinfo {year} {2008})}\BibitemShut {NoStop}%
\bibitem [{\citenamefont {J.Zeng}\ \emph {et~al.}(2015)\citenamefont {J.Zeng},
  \citenamefont {Li}, \citenamefont {Li}, \citenamefont {Dai}, \citenamefont
  {Tie},\ and\ \citenamefont {Lan}}]{p16}%
  \BibitemOpen
  \bibfield  {author} {\bibinfo {author} {\bibnamefont {J.Zeng}}, \bibinfo
  {author} {\bibfnamefont {J.}~\bibnamefont {Li}}, \bibinfo {author}
  {\bibfnamefont {H.}~\bibnamefont {Li}}, \bibinfo {author} {\bibfnamefont
  {Q.}~\bibnamefont {Dai}}, \bibinfo {author} {\bibfnamefont {S.}~\bibnamefont
  {Tie}}, \ and\ \bibinfo {author} {\bibfnamefont {S.}~\bibnamefont {Lan}},\
  }\href@noop {} {\bibfield  {journal} {\bibinfo  {journal} {Optics Express}\
  }\textbf {\bibinfo {volume} {23}},\ \bibinfo {pages} {31817} (\bibinfo {year}
  {2015})}\BibitemShut {NoStop}%
\end{thebibliography}%
\end{document}